\documentclass[useAMS,usenatbib,usegraphicx,referee]{mn2e}

\title[The scattering and extinction properties of nanodiamonds]{The scattering and extinction properties of nanodiamonds}

\author[Rakesh K Rai and Shantanu Rastogi]{Rakesh K Rai and Shantanu Rastogi\thanks{E-mail: shantanu\_r@hotmail.com}\\
Department of Physics, D.D.U. Gorakhpur University, Gorakhpur - 273009, India}

\begin{document}

\maketitle

\label{firstpage}

\begin{abstract}
The study of scattering and extinction properties of possible nanodiamond grains in the ISM are reported. Calculations using Discrete Dipole Approximation (DDA) for varying ellipsoidal shapes and sizes from $2.5$ to $10 ~nm$ are considered. Nanodiamonds show negligible extinction from IR to near-UV and very sharp far-UV rise. Comparison with observations rule out possibility of independent nanodiamond dust but point towards possibility of nanodiamonds as a component in the ISM. Radiation induced transformations may lead to carbonaceous grains with different core and mantles. So calculations are also performed for a core-mantle target model with nanodiamond core in graphite mantles. The graphite extinction features get modified with the peak at 2175 \AA{} being lowered, broadened, blue shifted and accompanied by enhanced extinction in the far-UV. Such variations in the 2175 \AA{} band and simultaneous far-UV rise are observed along some sources. A three component dust model incorporating silicate, graphite and graphite with nanodiamond core is also considered. The model extinction compares very well with the average galactic extinction in the complete range from $0.2$ to $10 ~\mu m^{-1}$. The best fit requires small size and small number of nanodiamonds.

\end{abstract}


\section{Introduction}

To account for the ultraviolet extinction in the diffuse interstellar medium \citet{saslaw69} first proposed the possibility of diamonds in Interstellar Medium (ISM). The 3.43 and 3.53 $\mu m $ emission bands in circumstellar media of Ae/Be Herbig stars HD~97048 and Elias~1 show convincing presence of nanodiamonds \citep{guillois99}. 
\citet{kerckhoven02} attribute these bands to C-H stretching in hydrogenated nanodiamond, as distinct from the 3.3 $\mu$m PAH feature. The 3.47 $\mu m $ feature in absorption toward a number of proto-stars is attributed to the tertiary C-H stretching mode in diamond like structures \citep{allamandola92}. Spectra of diamondoid molecules \citep{pirali07} show that nanodiamonds a few nanometre in size could be responsible for the 3.53 and 3.43 $\mu$m emission lines and smaller diamondoids give the 3.47 $\mu$m absorption feature. 

Cosmic nanodiamonds are also detected in primitive carbonaceous meteorites of presolar origin \citep{lewis87, daulton96, dai02, garai06}. In fact, nanodiamonds are considered to be the most abundant presolar component in meteorites \citep{anders93, zinner98}. \citet{jones04} studied $C-H$ stretching mode of nanodiamonds extracted from Orgueil meteorite and observed the above mentioned IR bands. If the ISM nanodiamonds are not hydrogenated their detection is hard \citep{kerckhoven02} and they could be more abundant than estimated by the observations of infrared $C-H$ bands \citep{dedeigo07}.

It is, therefore, important to incorporate nanodiamonds in UV extinction models for which their independent scattering and extinction properties need to be understood. The extinction properties of nanodiamond dust have been reported \citep{mutschke04} for spherical shape and the far-UV break of spectral energy in quasars is attributed to nanodiamond dust \citep{binette05, binette06}. In this communication extinction properties of nanodiamond grains of different ellipsoidal shapes and sizes are reported. To study the effect of nanodiamond within graphite extinction efficiencies are also reported for nanodiamond-graphite as core-mantle grains. An extinction curve modeling is proposed including nanodiamond as a component.

\section{Structured carbon metamorphs in ISM}

Various aspects of evolution of Carbon in the ISM have been discussed in literature \citep{dorschner95, henning98}. It is generally assumed that the ISM carbon is primarily Hydrogenated Amorphous Carbon (HAC) that transforms to graphite and diamond like structures in the presence of UV radiation field \citep{ligrn02}. Regions with low UV radiations, such as dense clouds and giant star atmospheres, graphitic particles may be absent \citep{sorrell90}. The estimated time-scale of graphitization varies from few weeks for circumstellar grains \citep{hecht86} to $\sim10^8 $ years for interstellar grains \citep{sorrell90}. Nanodiamonds can result from annealing of the carbonaceous material by UV radiation e.g. in Supernova ejecta \citep{nuth92}. Observations show that the hydrogenated nanodiamond features arise in the inner regions close to the central star, having higher UV flux \citep{goto09, kerckhoven02}. Surface energy studies show that small diamonds are more stable against thermal evaporation and chemical attack compared to graphite particles \citep{nuth87}.

Formation of nanodiamonds has been studied in laboratory giving insight to possible routes of their formation in ISM. Nanodiamond formation is seen to result from carbon vapour deposition (CVD) \citep{ugarte95, andersen98}, detonation of carbon based explosives \citep{krueger05}, electron irradiation of graphitic shells (carbon onions) \citep{banhart96, li08} and ion beam irradiation of amorphous carbon films \citep{sun99}. \citet{kouchi05} on the basis HRTEM studies suggest that diamonds and graphite few nanometre in size are formed by nucleation in organic ice mixture subjected to UV photolysis. Structural rearrangement between diamond and graphite is also possible. Graphite subjected to annealing/irradiation may go to shelled onion like form \citep{ugarte93} or transform into nanodiamond \citep{zaiser97}. Laboratory experiments also show that the $sp^3/sp^2 $ ratio changes by different doses of UV radiation \citep{ogmen88, duley-williams95}. Chemical conversion of PAH clusters to nanodiamonds is proposed by \citet{duley-grishko01}.

Partial graphitization of nanodiamond due to heat \citep{leguilou07} or pressure conditions \citep{davydov07} is suggested on the basis of some experimental studies. Theoretical analysis by \citet{kwon08} show that surface graphitization of nanodiamond under strong radiation field can lead to core-mantle like shell structure with upto 80\% graphitization. Thus nanodiamonds are possible within ISM carbonaceous matter and would modify the optical properties of dust. Graphitic mantle will make the nanodiamond crystals chemically less active and hard to detect spectroscopically.

\section{Calculation of Extinction Efficiencies} 

To calculate the extinction, scattering, absorption and polarization efficiencies the Discrete Dipole Approximation (DDA) is used \citep{purcell, draine88, draine94}. The DDSCAT 6.1 (http://www.astro.princeton.edu/draine/DDSCAT.htm) is utilized for the calculations. In DDA the continuous target is discretized by an array of dipoles placed at cubic lattices and the dielectric constant of the target material as a function of wavelength at ISM temperatures is required.

In general, optical properties of nano-sized particles are different from that of the bulk due to additional collisions of conducting electrons with the grain boundary. The imaginary part of the dielectric constant is changed to accommodate size effects and expressed as: $\varepsilon''= \varepsilon_{bulk}''+\frac{v_F \omega_{p}^{2}}{\beta a \omega^{3}} $ \citep{bohren, li04}, where $v_F$ is the electron velocity at the Fermi surface, $ \omega_p$ is the plasma frequency, $\beta$ is a constant of order unity and $\omega$ is the frequency of incident electromagnetic radiation. The second term increases with decreasing size and decreases rapidly with increase in frequency. As a result the dielectric constant does not vary appreciably with size in high frequency visible and UV regions\footnote{Considering incident radiation of 3000 \AA{} and the graphite $E_{\perp}$ data for $v_F$, $ \omega_p$ and $\beta$ from \citet{draine-lee84} the imaginary part of dielectric constant becomes $\varepsilon''= \varepsilon_{bulk}''+\frac{0.084}{a}$, where a is in nanometer. At this wavelength $\varepsilon_{bulk}''$ is $\sim$ 3.2 and will be negligibly affected by particle size (second term). The free electron density in diamond is smaller than graphite so the corresponding $v_F$ and $\omega_p$ will be smaller. Thus at high frequencies size affect in the case of nanodiamonds will be even smaller.} whereas, it varies slightly in the IR. The extinction, absorption and scattering efficiencies for nano-sized particles are very small in IR compared to visible and UV region, so small change in refractive index in the IR will not affect the result in general. Thus the refractive index data of bulk graphite and diamond are taken for all nano-sized particles.

Meteoritic nanodiamonds, with their density lower by 30\% and relatively high hydrogen content \citep{dorschner95}, differ from terrestrial diamonds. While terrestrial diamonds are cubic, the meteoritic diamonds are of Allende type \citep{lewis87} (small nanodiamonds found within Allende meteorites) or Carbonado diamonds \citep{garai06} (relatively large, black and porous). ISM nanodiamonds can be assumed to have properties more like the meteoritic nanodiamonds. For the present work we consider Allende nanodiamond to be the candidate material and take the dielectric constants reported by \citet{mutschke04}.

The size of meteoritic nanodiamonds are reported to be $\leq 5 ~nm$ in Allende type \citep{lewis87} and between $1$ to $3 ~nm$ in Orgueil meteorite \citep{daulton96}. Theoretical study of nanodiamonds with fullerene like surface \citep{raty03} suggests particle sizes from $1$ to $3 ~nm$. \citet{nuth87} report that graphite particles with radius $\approx 5 ~nm$ are thermodynamically unstable while the diamond structure is stable. IR spectroscopic study of diamondoids \citep{pirali07} points towards possibility of $\sim 10 ~nm$ nanodiamond in ISM while \citet{sheu02} suggest that the 3.43 and 3.53 $\mu m $ emission features are observable for nanodiamond $\geq 10 ~nm$. \citet{kerckhoven02} suggests log-normal particle size distribution between $1$ to $10 ~nm$ with peak diameter at $2.4 ~nm$. For the present work nanodiamond size is taken to be between $1$ to $10 ~nm$.

\section{Results and Discussion}

The target nanodiamonds are taken as oblate and prolate spheroids and ellipsoids and their different axial ratios as given in table 1. The size of these ellipsoids are taken such that the effective volume for each is equal to a sphere of radius $10$, $7.5$, $5.0$ and $2.5 ~nm$, so that nearly all possible nanodiamond sizes are taken into account. Table 1 also gives, for $10 ~nm$ nanodiamonds, the largest semi-axes and the number of dipoles used in DDA calculations. The validity criteria for DDA i.e. $|m|kd\leq 1$ is well considered and in none of the calculation it is more than $0.1$. To reduce calculation time 9 orientations for nanodiamond are considered as calculation for all 27 orientations done for some cases makes a difference of less than $0.1 \%$.

\begin{figure*}
\centering
\includegraphics[width=100 mm, angle= -90]{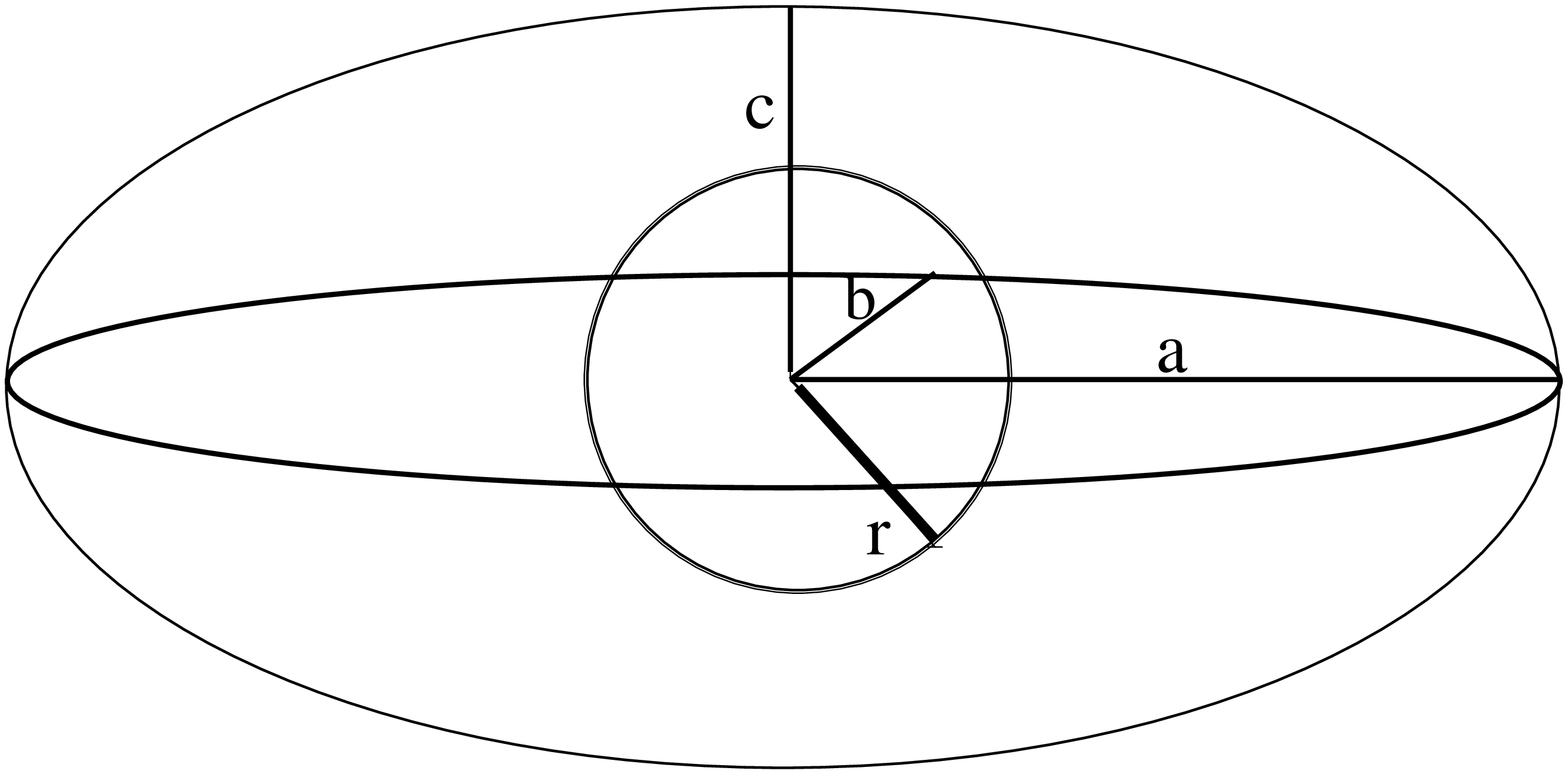}
\caption{Spherical nanodiamond core of radius `r' in graphitic ellipsoid of semi-axes `a', `b' and `c'}
\label{fig1}
\end{figure*}

The core-mantle model consisting of spherical nanodiamond core in different ellipsoidal graphitic mantle is represented in fig.\ref{fig1} with the ellipsoid semi axes `a', `b' and `c' and spherical core of radius `r'. The last column in table 1 gives nanodiamond percentage in total grain volume for `a' and `r' in the ratio $4:1$. All nanodiamond-graphite (core-mantle) calculations are done for 27 orientations. Calculations are done for 64 wavelengths from far-UV to near IR ($0.1 - 5.0 ~\mu m$) and linear in $\lambda ^{-1}$.

\begin{table}
 \centering

\caption{Ellipsoid axial ratios, largest axis, no. of dipoles considered \& Nanodiamond \% in core-mantle calculations}
  \begin{tabular}{c|c|c|c}

\hline
Ellipsoid   & Largest   & No. of dipoles &  Nanodiamond \% \\
axial ratio & semi-axis & used in        & for `a' and `r'\\
 a b c      & for $10~nm$ & calculation    & in ratio 4:1 \\
\hline
 4 4 3      & 11.006    &  43360         & 2.0830 \\

 4 4 2      & 12.599    &  29056         & 3.1250 \\

 4 4 1      & 15.874    &  14480         & 6.2500 \\

 4 3 3      & 12.114    &  32592         & 2.7777 \\

\textbf{~4 3 2~} & \textbf{13.867} & \textbf{21776} & \textbf{4.1666} \\

 4 3 1      & 17.472    & 10864          & 4.3333 \\

 4 2 2      & 15.874    & 14440          & 6.2500 \\

 4 2 1      & 20.000    & 7296           & 12.5000 \\

 4 1 1      & 25.198    & 3624           & 25.0000 \\
\hline

\end{tabular}
\end{table}

\subsection{Pure nanodiamonds} 

\begin{figure*}
\centering
\includegraphics[width= 150mm]{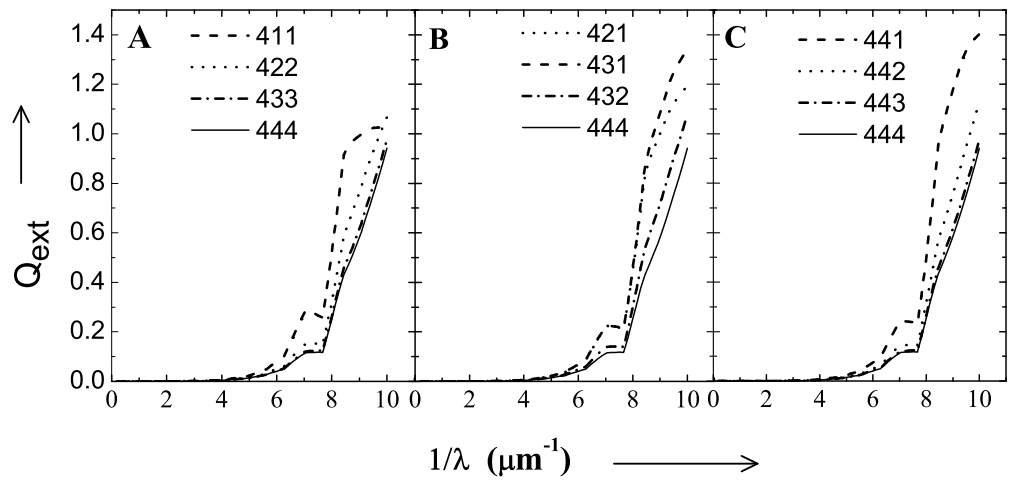}
\caption{Extinction efficiency for (A) Prolate spheroids, (B) Ellipsoids and (C) Oblate spheroids of effective radius $5 ~nm$ compared with that of nanodiamond sphere.}
\label{fig2}
\end{figure*}

The extinction efficiency variations for prolate spheroids, ellipsoids and oblate spheroids of effective radius $5 ~nm$ are plotted in fig.\ref{fig2}. The extinction efficiency for nanodiamonds of all shapes and the sizes is similar being negligible in the IR -- visible and increasing steeply in the UV. The extinction from 7.1 to 7.6 $\mu m^{-1}$ (1400 to 1300 \AA{}) becomes constant and even decreases slightly. This shape and size independent pause in extinction at 1400 \AA{} is also seen in figure 1 of \citet{binette05} for spherical nanodiamonds. The extinction efficiency is more for non-spherical shapes and the pause at 1400 \AA{} is more prominent for particles departing more from the spherical. \citet{mathis96} observed that extinction cross-section for spheroids are larger than those of spheres of same volume but \citet{voshchinnikov04} points out that this is true for small particles and in the forward direction only. This is also true for graphitic particles \citep{rgupta05}.

\begin{figure*}
\centering
\includegraphics[width= 150mm]{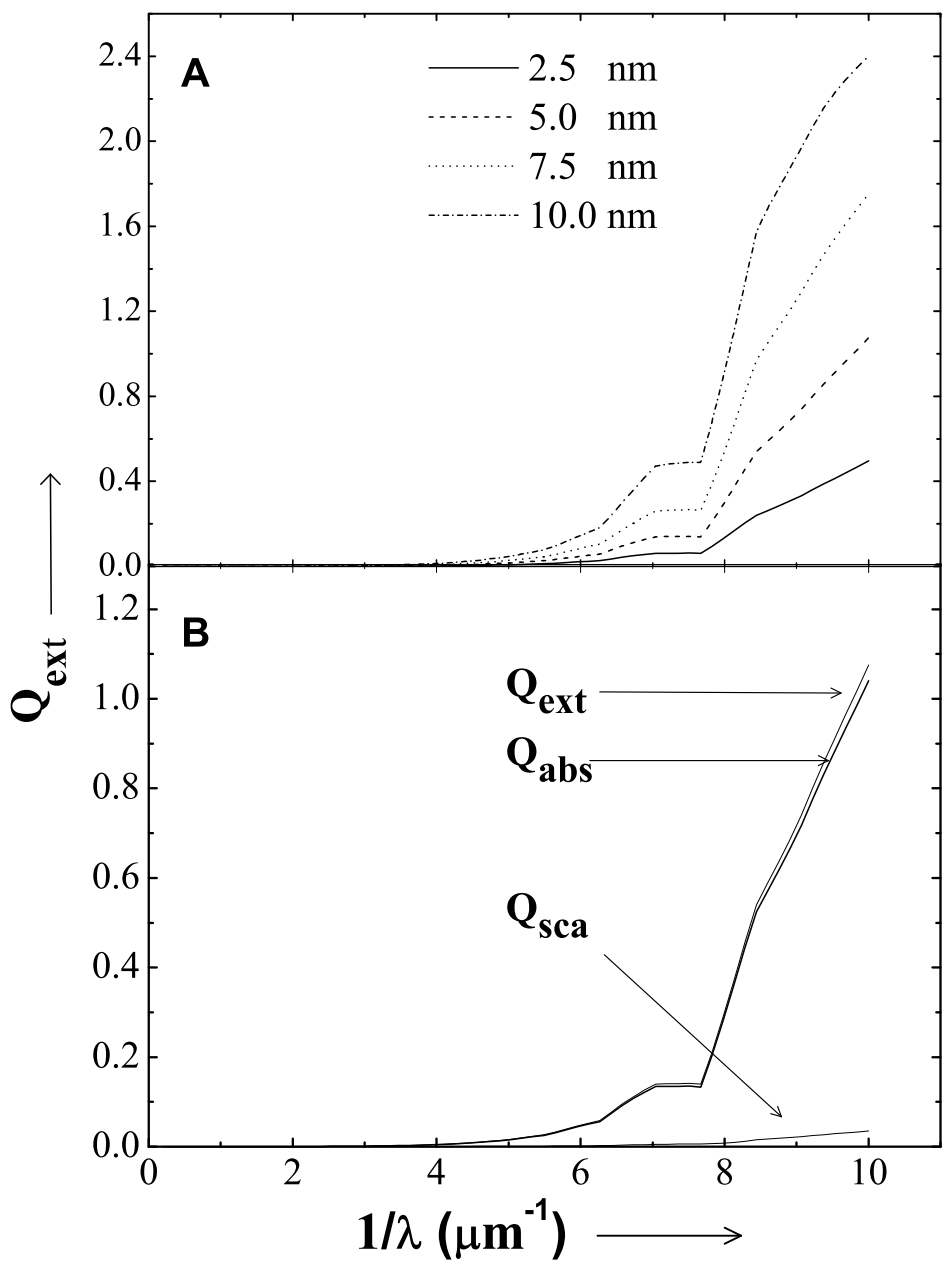}
\caption{(A) Extinction Efficiency for ellipsoid of shape 432 for different sizes, (B) Scattering and absorption efficiencies for effective size $5 ~nm$ }
 \label{fig3}
\end{figure*}

\begin{figure*}
\centering
 \includegraphics[width= 100mm]{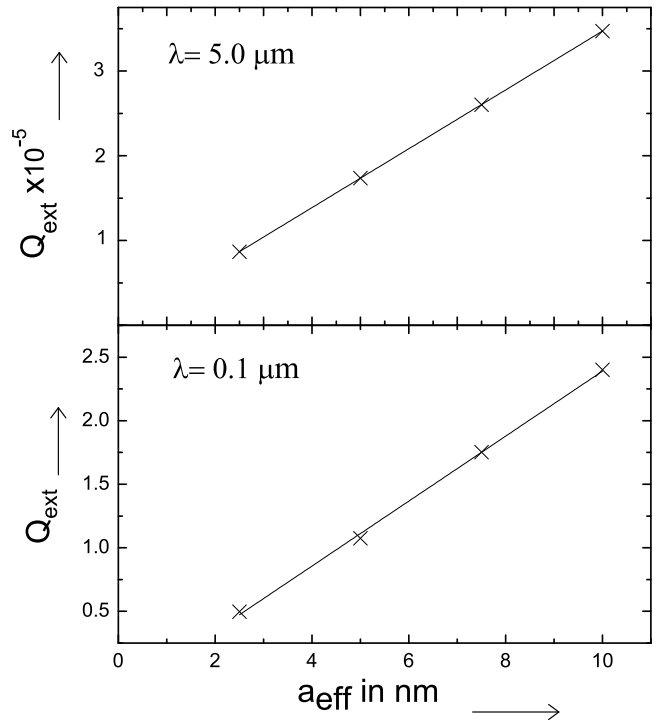}
\caption{Variation in $Q_{ext}$ with size for extreme wavelengths $\lambda = 5 ~\mu$m and $\lambda = 0.1 ~\mu$m}
  \label{fig4}
\end{figure*}

The extinction efficiency for ellipsoid of shape 432 for the four effective sizes is plotted in fig.\ref{fig3}(A), which shows increase in extinction with particle size. The scattering and absorption efficiencies for shape 432 of effective radius 5 nm is shown in fig.\ref{fig3}(B). The total extinction is almost all due to absorption even in profile. The scattering efficiency is much smaller in comparison. The extinction increases almost linearly with size for different wavelengths and is shown in fig.\ref{fig4} for two extreme wavelengths 5 $\mu $m (IR) and 0.1 $\mu $m (far-UV). This is true in general for nanosized particles \citet{draine93} and extinction from any in-between particle size can be extrapolated. 

\begin{figure*}
\centering
 \includegraphics[width=100mm]{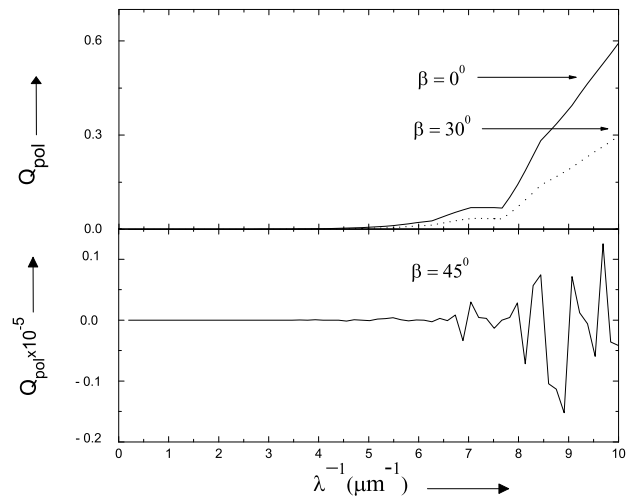}
\caption{Polarization efficiency for nanodiamond shape 432 and size $5 ~nm$ for orientation angles $\beta = 0, ~30^0 ~\& ~45^0 $ }
\label{fig5}
\end{figure*}

The polarization efficiency is defined as differential extinction cross-section in two orthogonal polarizations as $ Q_{pol}=\{Q_{ext}\parallel \textbf{e}\}- \{ Q_{ext}\perp \textbf{e} \} $ where $\textbf{e}$ is unit vector perpendicular to the direction of propagation. The polarization efficiency for shape 432 and size $5 ~nm$ is shown in fig.\ref{fig5}. $Q_{pol}$ follows the same trend as total extinction and decreases with increasing grain orientation angle $\beta$, as defined in DDSCAT \citep{draine04}. For $\beta = 45^0 $, $\{Q_{ext}\parallel \}$ and $\{Q_{ext}\perp \}$ are equal and polarization efficiency is nearly negligible. Small grains are hard to orient in a particular direction \citep{whittet}, thus nanodiamonds have negligible contribution in polarization.

In case of bulk diamond the energy band-gap $E_0$ is nearly 5.47 eV and for nanodiamonds an $a^{-2}$ gap size dependence gives $E_g = E_0 + 0.38(a/nm)^{-2}$ \citep{li04}. Variation in band-gap for nanodiamonds is observable only below $2~nm$ size \citep{raty03} and for smaller size diamondoids it is reported to be close to bulk diamond \citep{landt09}. The absorption of EM waves in nanodiamonds can be related to this band-gap as it starts from $\sim$ 4.7 $\mu m^{-1}$. The rise in extinction due to nanodiamond in far-UV region is abrupt and with increasing steepness. This large extinction explains the far-UV quasar break \citep{binette05} but the extreme far-UV extinction is not observed along any line of sight in the galaxy. The peak/plateau at 1400 \AA{} is also not observed in extinction profiles. Therefore, pure nanodiamonds are highly unlikely in ISM.

\subsection{Nanodiamond within graphite}

Using the optical properties given by \citet{lewis89}, \citet{aannestad95} incorporated nanodiamond in modeling the extinction curve and predicted a very low percentage of nanodiamond dust. Nanodiamonds may occur inside carbonaceous matter in ISM and manifest their presence by modifying the overall extinction profile. \citet{iati08} report triple layered grain with silicate, $sp^2 $ and $sp^3 $ carbonaceous material and \citet{yastrebov09} report nanodiamond enveloped in glassy carbon shells. Considering that surface graphitization of nanodiamond leads to core-mantle like shell structure \citep{kwon08, li08}, extinction efficiency calculations are reported for spherical nanodiamond inside graphite ellipsoidal mantle. 

Extinction properties of graphitic particles are well studied \citep{draine-lee84,draine93}. The $(2/3 - 1/3)$ approximation for dielectric anisotropy is found to be good for small particles and can also be used for graphite with coatings \citep{draine93}. Refractive index data of $10 ~nm$ graphite at 20 K is taken from B. T. Draine's website and the $(2/3 - 1/3)$ approximation applied.

\begin{figure*}
\centering
 \includegraphics{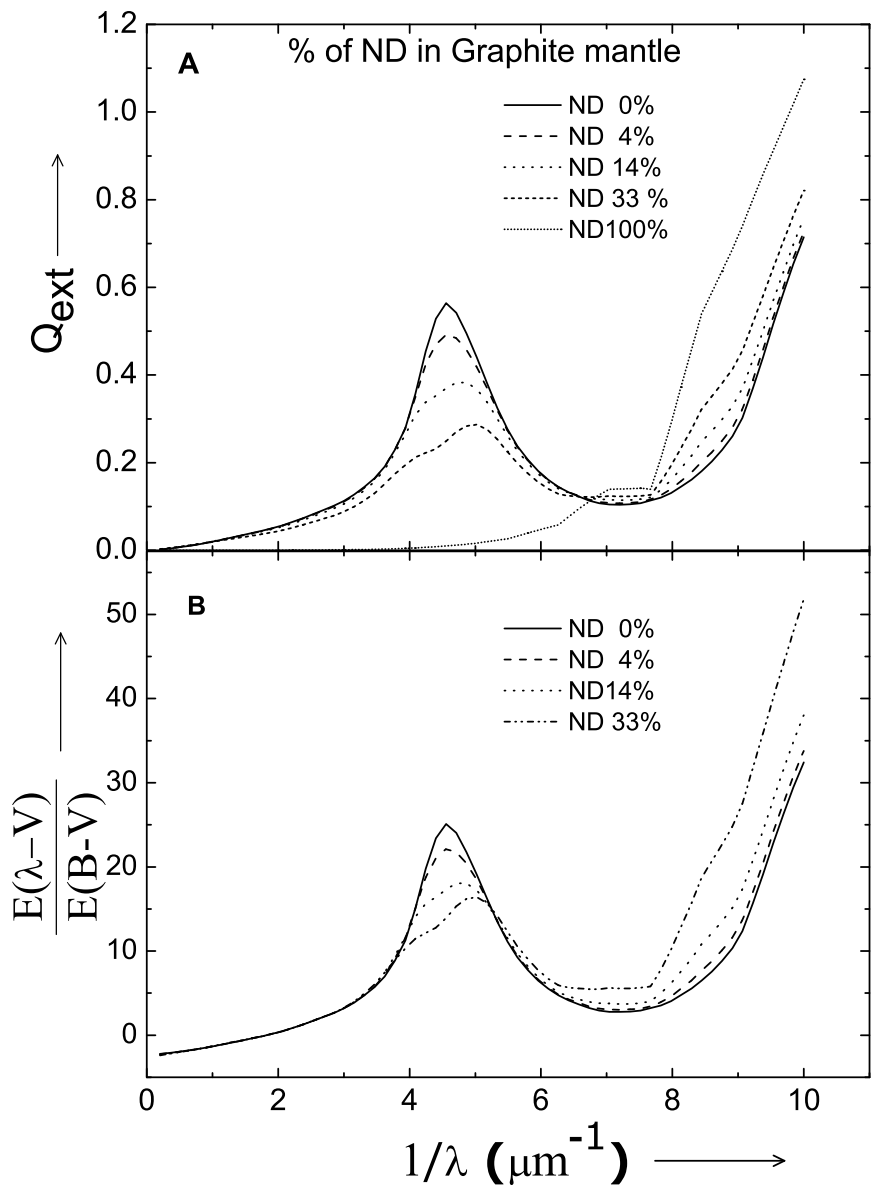}
\caption{(A) Extinction efficiency and (B) Normalized extinction for different nanodiamond percentage in graphite mantle of shape 432 and effective radius $5 ~nm$}
\label{fig6}
\end{figure*}

The nanodiamond-graphite core-mantle extinction efficiency and normalized extinction are shown in fig.\ref{fig6}, for mantle shape 432 and size $5 ~nm$. The volume percentage of core nanodiamond is also mentioned. It is seen that the 2175 \AA{} peak in graphite gets modified due to the presence of nanodiamond core. On increasing the nanodiamond percentage, the peak is lowered, broadened and slightly blue shifted. In the far-UV region there is sharp rise in extinction that is steeper for larger nanodiamond percentage. Observed extinction along various lines of sight show similar 2175 \AA{} feature lowering and broadening associated with enhanced far-UV rise \citep{cardelli88, fitz-massa90, fitz-massa05, fitz-massa07}. Some such stellar regions, viz. HD210121, HD204827, HD29647, that do not follow the CCM rule \citep{CCM89} are termed as non-CCM sites \citep{valencic04}. Also there are some sight-lines, e.g. HD3191, HD284839, HD284841, HD287150 etc. \citep{fitz-massa07}, that exhibit similar broad bump and steep far-UV rise.

\begin{figure*}
\centering
 \includegraphics[width= 150mm]{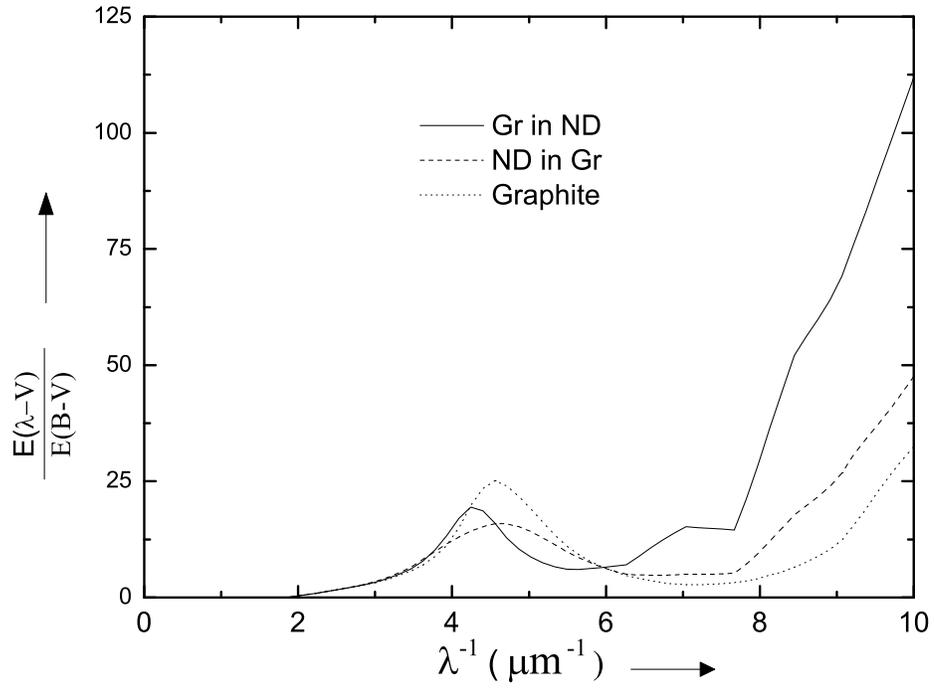}
\caption{Extinction for pure graphite, nanodiamond in graphite mantle and  graphite core in nanodiamond mantle; shape 432, mantle size $5 ~nm$ and core 33 \% by volume}
\label{fig7}
\end{figure*}

Incorporating a mantle broadens the 2175 \AA{} peak and to model this feature \citet{mathis94} considered different mantle material including diamond over graphite. Taking similar target with Allende nanodiamond mantle and graphite core normalized extinction is plotted in fig.\ref{fig7}, wherein the values for pure graphite and nanodiamond core in graphite mantle are also shown for comparison. The core is taken to be $33$\% by volume. With both mantles the 2175 \AA{} peak is lowered but for graphite core it shifts towards visible as opposed to observations \citep{CCM89,valencic04}. The peak/plateau between 1300 to 1400 \AA{} of pure nanodiamond is prominent in the case of nanodiamond mantle while it smooths out when nanodiamond is taken as core. Similarity of the extinction profile of nanodiamond core--graphite mantle with observations strengthens the possibility of surface graphitization of $sp^3$ diamond structures in UV dominated regions and also explains the absence of diamond $C-H$ stretching features along most sight-lines \citep{acke06}.

\subsection{Modeling galactic extinction}

To understand features of the extinction curve carbonaceous species, especially graphite, and silicates in several shapes, sizes and coatings have been the main components of most extinction models \citep{whittet, voshchinnikov04}. The far-UV rise in extinction has been largely unexplained though very small grains of graphite and silicates \citep{hong80, sorrell89, ligrn97} and polycyclic aromatic hydrocarbon (PAH) molecules \citep{drnli01,lidrn01} have been considered. \citet{zubko04} used a wide variety of carbon and silicate materials in different proportion with fourteen parameter size distribution function to reproduce the extinction curve while \citet{cecchi08} consider a mixture of neutral, cation, dication PAHs.

Most observations show relationship between lowering of the 2175 \AA{} feature with rise in far-UV extinction \citep{cardelli88, valencic04, clayton06}. This simultaneous modification of the 2175 \AA{} feature and far-UV rise is indicated in our study incorporating nanodiamond core within graphite. We present a model for the mean galactic extinction considering silicates, graphite and nanodiamond core within graphite mantle as the three components of ISM dust. For a simple model spherical silicate and graphite grains are considered. Mie theory (BHMIE code from \citet{bohren}) is utilized in calculations at 30 wavelengths and for each wavelength 50 radii are used in \citet{mrn77} distribution. As the third component nanodiamond-graphite (core-mantle) is considered. Ellipsoid mantle with axial ratio 432 and spherical nanodiamond core in three different volume percentages as shown in Table 2 are taken. No particle size distribution is chosen, as $Q_{ext}/a$ is a constant and the normalized extinction will be essentially the same for all the particle size distributions.

\begin{figure*}
\centering
 \includegraphics[width=140mm]{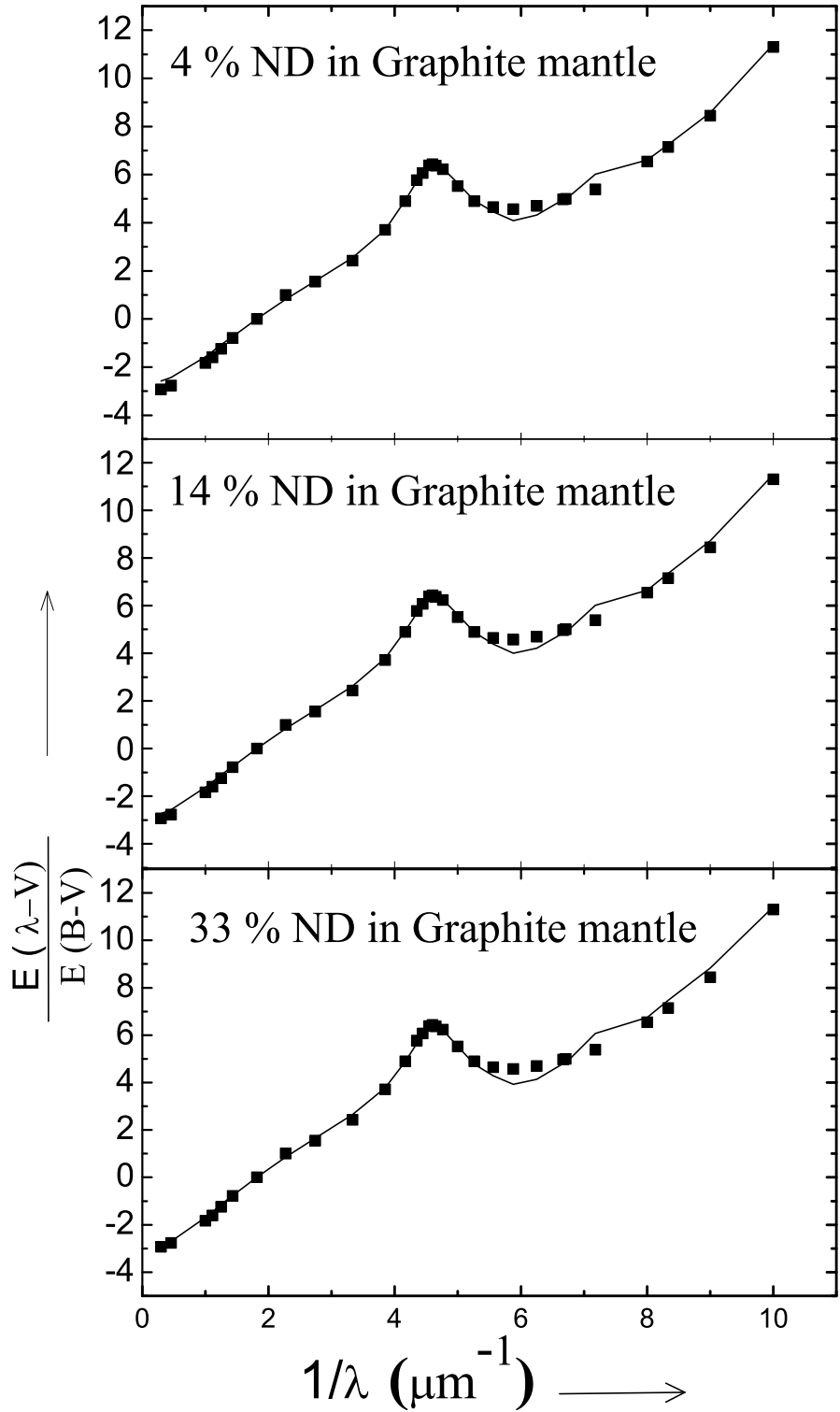}
\caption{The three component modelling of extinction curve consisting of silicate, graphite and nanodiamond-graphite (core-mantle) with \% of nanodiamond in graphite mantle is shown as label}
\label{fig8}
\end{figure*}

The three components are linearly combined with p, q and s contributory weights and resulting extinction is compared with the average galactic values \citep{whittet}. The set of reduced $ \chi^2_j$ \citep{bevington69,vaidya01} are given by
\begin{displaymath}
 \chi^{2}=\frac{\sum^n_{i=1} (S^j_i-T_i)^2}{pp}
\end{displaymath}
where pp is the degrees of freedom, $S^j_i(\lambda_i)$ is the j\textit{th} model curve (j=1 to 3 for silicate, graphite and nanodiamond-graphite p, q and s fractions respectively) and $T_i(\lambda_i)$ is the observed value at wavelength $\lambda_i$. The p, q, s combinations that give minimized $\chi^2$ are shown in Table 2 with corresponding graphs in fig.\ref{fig8}. The 2175 \AA{} peak modifications and the far-UV rise are simultaneously explained on incorporation of very small percentage of nanodiamonds.

\begin{table}
 \centering
\caption{Contributory weights of the three components, $\chi^2$ values and different nanodiamond percentage in graphite}
\begin{tabular}{@{}r|c|c|c|c|c@{}}
\hline
      &  p & q & s & $\chi^2$ & Core nanodiamond size \\
      &    &   &   &          &  in 5 nm grain \\
\hline
4  \% &  0.40  &  0.42  &  0.03 &  0.032 & 1.7 \\
14 \% &  0.39  &  0.43  &  0.03 &  0.038 & 2.6 \\
33 \% &  0.39  &  0.47  &  0.01 &  0.048 & 3.4 \\
0  \% &  0.40  &  0.49  &  --   &  0.052 & --  \\
\hline

\end{tabular}
\end{table}

For comparison Table 2 also shows the $0\%$ nanodiamond, two component model, that gives 0.40 silicate and 0.49 graphite fractions. This is well within estimated silicate and graphite grain abundances \citep{draine-lee84}. On adding nanodiamond-graphite core-mantle particles as the third component the fit improves. This very small grain component is a small fraction of carbonaceous particles having even smaller nanodiamond core. That is much lower than the $10\%$ upper limit of ISM nanodiamond reported by \citet{lewis89}. The core-mantle component also reduces the graphite fraction slightly and changes negligibly the relative fraction of silicates. The $\chi^2$ values increase with increasing percentage of nanodiamond in graphite mantle but its overall contribution (fraction `s') decreases. The fraction of nanodiamond-graphite is also much smaller than used by \citet{aannestad95} for $5~nm$ nanodiamond grains. Considering nanodiamond density of 3.52~gm/cc the nanodiamonds in our models are $< 0.1\%$ fraction of carbonaceous particles. The best fit, in the complete 0.2 to 10 $\mu m^{-1}$ wavelength range, stands for smaller nanodiamond in larger amount. For effective grain size of $5~nm$, the $4 \%$ nanodiamond core implies $1.7~nm$ size, which is about the size of nanodiamond in meteorites \citep{lewis87, daulton96}.

The above results assume only Allende type nanodiamonds in ISM. Alternatively ISM nanodiamonds could be like terrestrial cubic diamonds, sp3 hybridized polymeric carbon or a complex mixture of these forms. Different types of diamonds have different dielectric functions and will lead to slightly different extinction profiles. While \citet{iati08} consider sp3 polymeric carbon coatings, \citet{binette05, binette06} consider both cubic and Allende type nanodiamonds. Allende type nanodiamonds show far-UV steepening and peaking at lower frequencies than cubic diamonds and the far-UV break in quasars is better represented when greater proportion of Allende type nanodiamonds are considered \citep{binette05, binette06}.

\section{Conclusions} 

Various experimental, theoretical and observational studies suggest the possibility of carbon in diamond form in the ISM. Besides enhanced far-UV extinction \citep{binette05, binette06}, luminescence from diamond nano-crystals could be responsible for the Extended Red Emission (ERE) \citep{chang06}. The study of extinction properties of possible nanodiamond grains of different shapes and sizes is done to understand their role in far-UV extinction and in modification of the overall extinction curve. 

The extinction due to nanodiamonds of all shapes and sizes is essentially similar with negligible extinction from IR to near UV range and sharp rise in extinction in the far-UV. The extinction due to non-spherical shape is higher than that due to spherical. For nanosized particles the extinction efficiency increases linearly with effective radius.

Considering nanodiamond-graphite core-mantle target, modification in the 2175 \AA{} peak and graphite extinction is studied. In general the 2175 \AA{} peak gets lowered, broadened and there is enhanced far-UV extinction. Increase in \% of $ sp^3 $ character in graphite gives rise to higher far-UV extinction. The three component modeling of average galactic extinction gives best results for small sized and low \% of nanodiamond. This enables us to explain far-UV rise and simultaneous modification of the 2175 \AA{} bump without putting constraints on the interstellar abundances.

Extinction from specific objects that show enhanced far-UV rise can be attempted by incorporating a nanodiamond component. This will enhance understanding of radiation induced transformations of carbonaceous matter in the ISM.

\section*{Acknowledgements}
The use of High Performance Computing and library facilities at IUCAA, Pune is acknowledged. Valuable discussions with Dr. R. Gupta and Dr. D.B. Vaidya are greatly appreciated.

\newpage

\newpage

\label{lastpage}

\end{document}